# The Expanding Universe, Planetary Motion and The Pioneer 10/11 Anomaly.

A. Lewis Licht
Dept. of Physics
U. of Illinois at Chicago
licht@uic.edu

**Abstract**

The effect of the expanding universe on planetary motion is considered to first order in the Hubble constant H. The orbital elements are shown to be unaffected, that is, the orbits do not expand, and have indeed no extra secular variation. However, there is a small periodic change in the connection between planetary proper time and coordinate time. This has the effect of producing an apparent anomalous acceleration in velocities inferred from echo-ranging, but the effect is too small by many orders of magnitude to account for the Pioneer 10/11 anomaly.

## 1 Introduction

The detection of an anomalous acceleration in the Pioneer 10/11 and Ulysses trajectories [1] [2] has led to the suggestion that this acceleration might have a cosmological cause [3]. It was pointed out by Rosales and Sanchez-Gomez [3] that, with H denoting the Hubble constant, and with H = 85 km/sMps, the combination

$$Hc = 8.5 \times 10^{-10} \text{ m/s} \qquad (1)$$

was exactly equal to the magnitude of the anomalous acceleration. They did not prove that this was the acceleration, but they did derive an expression for the light time that contained an extra term proportional to the square of the space probe's coordinate time. However Turyshev et al [2] found that although a term of this form could account for the Pioneer 10/11 and Galileo anomalous acceleration, it would have the wrong sign for Ulysses. These considerations raise the question of what exactly is the effect of cosmic expansion on planetary motion. We attempt to answer this question here.

In the following we derive, in Newtonian approximation, to first order in H, the metric about a gravitating point mass in the presence of a cosmological expansion and find the effect of the expansion on planetary orbital elements. We then do an operational analysis of the process of echo-ranging and find an extremely small anomalous acceleration, too small in fact to account for the Pioneer 10/11 anomaly. We use units where c = G = 1.

## 2 The metric

The Friedman-Walker metric for an expanding flat space [4] is

$$ds^2 = -dt^2 + a(t)^2[dr^2 + r^2 d\Omega^2] \tag{2}$$

Where a(t) is the expansion scale factor. We denote $H = \frac{\dot{a}}{a}$. This metric is a solution to the Einstein equations:

$$G_{\mu\nu} = \frac{8\pi\rho}{3} u_\mu u_\nu \tag{3}$$

for a dust filled universe with uniform density $\rho$, where $u^\mu = (1, \mathbf{0})$, the local dust four-velocity.

In the presence of a gravitating point mass, in Newtonian approximation the metric of a static asymptotically flat space is:

$$ds^2 = -\left[1 - \frac{2m}{r}\right]dt^2 + \left[1 + \frac{2m}{r}\right]dr^2 + r^2 d\Omega^2 \tag{4}$$

To first order in m, H, and Hm, a solution to the problem of the metric about a point mass in an expanding universe is given by:

$$ds^2 = -\left[1 - \frac{2m}{ar}\right]dt^2 + a^2\left\{\left[1 + \frac{2m}{ar}\right]dr^2 + r^2 d\Omega^2\right\} \tag{5}$$

where a is the expansion scale factor of Eq. (2).

The proof of this assertion consists in using the metric of Eq. (5) to compute G to lowest order. The additional terms in Eq. (5) only produce high order perturbations in $G_{\mu\nu}$. Namely,

$$G_{tt} - \frac{8\pi}{3}\rho \sim O(mH^2) \quad , \quad G_{rr} \sim O(mH^2) \quad , \quad G_{\theta\theta} \sim O(H^2) \quad , \quad G_{rt} \sim O(Hm^2) \tag{6}$$

We change now to a coordinate system where

$$u = ar \, , \tag{7}$$

then

$$ds^2 = -\left[1 - \frac{2m}{u} - H^2 u^2\left(1 + \frac{2m}{u}\right)\right]dt^2 - 2H(u + 2m)dtdu + \left[1 + \frac{2m}{u}\right]du^2 + u^2 d\Omega^2 \, , \text{ or}$$

$$ds^2 = -\left[1 - \frac{2m}{u}\right]dt^2 - 2H(u + 2m)dtdu + \left[1 + \frac{2m}{u}\right]du^2 + u^2 d\Omega^2 \tag{8}$$

to lowest order.

It turns out that the (u,t) system is more appropriate for planetary analysis. We will prove that the cosmological expansion has, to this order, no effect on planetary orbital parameters. That is, the orbits do not expand. There is however a small effect on the relation between coordinate and planetary proper time, which might have an effect on ranging.

The (t, u) part of the metric tensor is:

$$g = \begin{pmatrix} -1 + \frac{2m}{u} & -H(u + 2m) \\ -H(u + 2m) & 1 + \frac{2m}{u} \end{pmatrix} \tag{9}$$

with inverse:

$$g^{-1} = \begin{pmatrix} -1 - \frac{2m}{u} & -H(u + 2m) \\ -H(u + 2m) & 1 - \frac{2m}{u} \end{pmatrix} \tag{10}$$

The Christoffel symbols in this metric are

$$\Gamma^u_{00} = \frac{m}{u^2} \, , \quad \Gamma^u_{0u} = \frac{Hm}{u} \, , \quad \Gamma^u_{uu} = -\frac{m}{u^2} \, , \quad \Gamma^0_{00} = -\frac{Hm}{u} \tag{11}$$

$$\Gamma^0_{0u} = \frac{m}{u^2} \, , \quad \Gamma^0_{uu} = H\left(1 + \frac{3m}{u}\right) \, , \quad \Gamma^u_{u\theta} = 0 \, , \quad \Gamma^u_{\theta\theta} = -(u - 2m)$$

$$\Gamma^\theta_{\theta u} = \frac{1}{u} \, , \quad \Gamma^u_{\phi\phi} = -(u - 2m)\sin^2(\theta) \, , \quad \Gamma^\phi_{\phi u} = \frac{1}{u} \, , \quad \Gamma^\theta_{\phi\phi} = -\sin(\theta)\cos(\theta)$$

$$\Gamma^\phi_{\phi\theta} = \cot(\theta)$$

## 3   Planetary Motion

Let a dot denote $d/d\tau$, where $\tau$ denotes proper time. The planetary four-vector velocity then satisfies:

$$1 = \left[1 - \frac{2m}{u}\right]\dot{t}^2 + 2H(u + 2m)\dot{t}\dot{u} - \left[1 + \frac{2m}{u}\right]\dot{u}^2 - u^2\dot{\theta}^2 - u^2\sin^2(\theta)\dot{\phi}^2 \tag{12}$$

We need consider only motion in the u-$\theta$ plane. The equations of motion are then:

$$\ddot{u} + \Gamma^u_{00}\dot{t}^2 + 2\Gamma^u_{0u}\dot{t}\dot{u} + \Gamma^u_{uu}\dot{u}^2 + \Gamma^u_{\theta\theta}\dot{\theta}^2 = 0 \tag{13}$$

$$\ddot{\theta} + 2\Gamma^\theta_{\theta u}\dot{\theta}\dot{u} = 0 \tag{14}$$

Or,

$$\ddot{u} + \frac{m}{u^2}\dot{t}^2 + 2\frac{Hm}{u}\dot{t}\dot{u} - \frac{m}{u^2}\dot{u}^2 - (u - 2m)\dot{\theta}^2 = 0 \tag{15}$$

$$\ddot{\theta} + \frac{2}{u}\dot{\theta}\dot{u} = 0 \tag{16}$$

From Eq. (16) follows the conservation of angular momentum:

$$u^2\dot{\theta} = \text{constant} \tag{17}$$

From Eq. (12) we find, to first order in H,

$$\dot{t} = 1 + \frac{m}{u} + \left(1 + \frac{3m}{u}\right)\frac{\dot{u}^2}{2} + \left(1 + \frac{m}{u}\right)\frac{u^2\dot{\theta}^2}{2} - H(u + 4m)\dot{u} \tag{18}$$

Eq. (15) now becomes, to order mH,

$$\ddot{u} + \frac{m}{u^2} - (u - 3m)\dot{\theta}^2 = 0 \tag{19}$$

The Hubble constant H does not appear in Eq. (19), therefore the orbital parameters are not affected by the cosmic expansion.

Let $t_{NC}$ denote coordinate time calculated as a function of proper time without the H term. We set $t - t_{NC}$ to be zero at perihelion, where $u = q$, the perihelion distance. Then from Eq. (18) the deviation in coordinate time is

$$t - t_{NC} = -\frac{H}{2}(u^2 - q^2) \qquad (20)$$

For an elliptical orbit, this is periodic, and there is no secular deviation between the two times. It is greatest at aphelion. For the Earth, it is at most $10^{-13}$ s.

## 4    Echo-Ranging

The orbital positions calculated from Eqs. (17) and (19) are functions of the planet's proper time. The difference between proper time and coordinate time can lead to anomalies in the comparision of the positions of different orbiting bodies. We illustrate this with the problem of determining from Earth the position of a space probe through echo ranging. For simplicity, we neglect terms of order m, and consider only the effect of the cosmological term H. The Earth's orbit is taken as circular, and the probe's trajectory is approximated as radial, with constant velocity.  In this approximation, the probe's proper time is

$$d\tau = \frac{du}{\dot{u}} \qquad (21)$$

and Eq. (8) becomes:

$$dt^2 + 2Hududt - du^2 = \frac{du^2}{\dot{u}^2} \qquad (22)$$

With the usual notation for the probe's spatial velocity, $v = \beta c$,

$$\dot{u} = \beta\gamma, \quad \gamma = 1/\sqrt{1 - \beta^2} \qquad (23)$$

Eq. (22) can be solved to first order in H as:

$$dt = \frac{du}{\beta} - Hudu \qquad (24)$$

resulting in

$$t_f - t_i = \frac{R_f - R_i}{\beta} - \frac{H}{2}\left[R_f^2 - R_i^2\right] \qquad (25)$$

for the change from position $R_i$ at time $t_i$, to $R_f$ at $t_f$. This may be solved for $R_f$ in lowest order as:

$$R_f = R_i + \beta(t_f - t_i) + \frac{\beta^2 H(t_f - t_i)}{2}(\beta(t_f - t_i) + 2R_i) \qquad (26)$$

from which it is clear that:

$$\left(\frac{dR_f}{dt}\right)_i = \beta + \beta^2 H R_i \quad , \quad \left(\frac{d^2 R_f}{dt^2}\right)_i = \beta^3 H \qquad (27)$$

Figure 1a shows a space probe moving radially outwards from the sun as seen from the North side of the Earth's Orbit. A polar axis is taken as the projection of the probe's trajectory into the plane of the ecliptic. At coordinate time $t_0$, when the Earth is at the position vector $\mathbf{R}_{E0}$, making an angle of $\theta_0$ with this axis, and the probe is at the position vector $\mathbf{R}_0$, a radio pulse is emitted from the Earth. It arrives at the probe when the probe is at the position vector $\mathbf{R}_1$, and is reflected back to Earth, arriving there at coordinate time $t_2$, when the Earth would be at the position vector $\mathbf{R}_{E2}$ and angle $\theta_2$ (not shown). Figure 1b shows a side view of the the situation. The probe's trajectory makes the angle $\alpha$ with the ecliptic plane.

Light travels along a zero geodesic, therefore:

$$dt^2 + 2Hudu\,dt - du^2 - u^2 d\Omega^2 = 0 \qquad (28)$$

Which may be solved to first order in H as:

$$dt = ds - Hudu \quad , \qquad (29)$$

where

$$ds = \sqrt{du^2 + u^2 d\Omega^2} \qquad (30)$$

If the light covers the spatial distance $\Delta s$, as the radius changes from $u_i$ to $u_f$, the coordinate time changes from $t_i$ to $t_f$, where, from Eq. (29):

$$t_f - t_i = \Delta s - \frac{H}{2}\left(u_f^2 - u_i^2\right) \qquad (31)$$

Applying this to the situation shown in Figure 1a, we have:

$$t_1 = t_0 + |\mathbf{R}_1 - \mathbf{R}_{E0}| - \frac{H}{2}\left(R_1^2 - R_E^2\right) \qquad (32)$$

This is essentially the result of Rosalez and Sanchez-Gomez. The H term looks as if it could provide an anomalous acceleration, however it cancels out when one adds in the time for the return trip. That is:

$$t_2 = t_1 + |\mathbf{R}_1 - \mathbf{R}_{E2}| - \frac{H}{2}\left(R_E^2 - R_1^2\right) \qquad (33)$$

So that the total elapsed coordinate time is:

$$t_2 - t_0 = |\mathbf{R}_1 - \mathbf{R}_{E0}| + |\mathbf{R}_1 - \mathbf{R}_{E2}| \qquad (34)$$

There is no explicit dependence on H in the elapsed total time. However there is still an implicit dependence. Knowing $\theta_2$ and $\theta_0$, $R_1$ may be determined from the measured time difference. We will see that there is a part of $\frac{d^2 R_1}{dt_0^2}$ that is proportional to H, the anomalous acceleration, $\delta \ddot{R}$. From Eq.(26) we have

$$R_1 = R_0 + \beta(t_1 - t_0) + \frac{\beta^2 H(t_1 - t_0)}{2}(\beta(t_1 - t_0) + 2R_0) \qquad (35)$$

and, applying Eq. (32) we get:

$$R_1 - R_0 - \frac{\beta H}{2}\left[R_1^2 - R_0^2\right] = \beta(t_1 - t_0) = \beta|\mathbf{R}_1 - \mathbf{R}_{E0}| - \frac{\beta H}{2}\left(R_1^2 - R_E^2\right) \qquad (36)$$

Which allows us to write:

$$R_1 - R_0 = \beta |\mathbf{R}_1 - \mathbf{R}_{E0}| - \frac{\beta H}{2}(R_0^2 - R_E^2) \tag{37}$$

Expanding to second order in $R_1 - R_0$:

$$R_1 - R_0 = \beta\left[\Delta s_0 + \frac{(R_0 - R_E\cos(\psi))}{\Delta s_0}(R_1 - R_0) + \frac{R_E^2\sin^2(\psi)}{2\Delta s_0^3}(R_1 - R_0)^2\right] - \frac{\beta H}{2}(R_0^2 - R_E^2) \tag{38}$$

where

$$\cos(\psi) = \hat{\mathbf{R}}_1 \cdot \hat{\mathbf{R}}_{E0} = \cos(\alpha)\cos(\theta_0) \quad , \quad \Delta s_0 = |\mathbf{R}_0 - \mathbf{R}_{E0}| \tag{39}$$

we can solve for $R_1$ in lowest order as:

$$R_1 = R_0 + \beta|\mathbf{R}_0 - \mathbf{R}_{E0}| + \beta^2(R_0 - R_E\cos(\psi)) + \frac{\beta^3 R_E^2 \sin^2(\psi)}{2|\mathbf{R}_0 - \mathbf{R}_{E0}|} - \frac{\beta H}{2}(R_0^2 - R_E^2) \tag{40}$$

From Eq.(27) we have:

$$\frac{dR_0}{dt_0} = \beta + \beta^2 H R_0 \quad , \quad \frac{d^2 R_0}{dt_0^2} = \beta^3 H \tag{41}$$

We will use this to differentiate (40), keeping only the terms proportional to H:

Now to second order in $R_E$

$$|\mathbf{R}_0 - \mathbf{R}_{E0}| = R_0 - R_E\cos(\psi) + \frac{R_E^2}{2R_0}(1 - \cos^2(\psi)) \tag{42}$$

and to lowest order in $\beta$ and $R_E/R_0$:

$$R_1 = R_0(1 + \beta) - \beta R_E\cos(\psi) + \beta\frac{R_E^2}{2R_0}(1 - \cos^2(\psi)) - \frac{\beta H}{2}(R_0^2 - R_E^2) \tag{43}$$

*8*

Using Eq. (41) we get for the anomalous acceleration

$$\delta\ddot{R}_1 = (\beta^3 + \beta^4)H + \beta\frac{R_E^2}{2}\left[\left[-\frac{\beta^3 H}{R_0^2} + 4\frac{\beta^3 H R_0}{R_0^3}\right](1 - \cos^2(\psi)) - \frac{2\beta^3 H R_0 R_E^2}{R_0^2}\cos^2(\alpha)\cos(\theta_0)\sin(\theta_0)\theta_0\right] - \frac{\beta H}{2}(2\beta^2) \quad (44)$$

$$= \beta^3 H\left[\frac{3}{2}\left(\frac{R_E}{R_0}\right)^2(1 - \cos^2(\alpha)\cos^2(\theta_0)) - \beta_E\left(\frac{R_E}{R_0}\right)\cos^2(\alpha)\cos(\theta_0)\sin(\theta_0)\right]$$

which is on the order of $10^{-33}$ m/s$^2$, much too small to account for the Pioneer 10/11 anomaly.

## 5  Conclusion

We have found the equations of motion in terms of proper time to first order in m, H and mH for planetary motion about a gravitating point mass in the presence of an expanding universe. The equations have no explicit dependence on H, therefore the orbital elements are unaffected by the expansion. However, there is a term involving H in the relation between proper time and coordinate time. As a result of this, measurement of solar system distances through echo-ranging can show an extremely weak H-dependence.

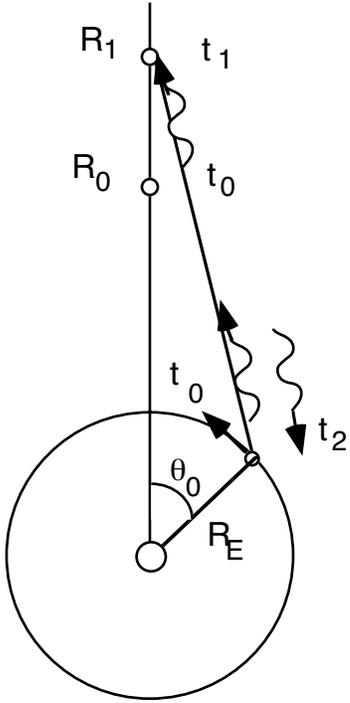
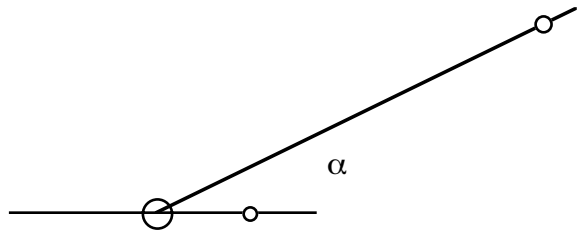

Figure 1a            Figure 1b

Figure 1:  The Echo-ranging geometry for a radially moving space probe.